\documentclass[aps,prd,amsmath,floats,floatfix, twocolumn,superscriptaddress,nofootinbib,showpacs,reprint]{revtex4}

\usepackage{hyperref}

\usepackage{graphicx}
\usepackage{xspace}
\usepackage[usenames,dvipsnames]{color}
\usepackage{amsmath,amssymb}
\usepackage{longtable}
\usepackage{ulem} 

\newcommand{\Cambridge}{\affiliation{DAMTP,University of Cambridge,
    Wilberforce Rd., Cambridge, CB3 0WA, England}}
\newcommand{\UNC}{\affiliation{Department of Physics and Astronomy,
    UNC-Chapel Hill, NC 27599, USA}}

\begin{document}

\title{
 Back-reaction of the Hawking radiation flux in Unruh's vacuum on a gravitationally collapsing star II}

\author{Laura Mersini-Houghton} \UNC \Cambridge 
\date{\today}

\begin{abstract}

  A star collapsing gravitationally into a black hole emits a flux of
  radiation, known as Hawking radiation. When the initial state of a
  quantum field on the background of the star, is placed in the Unruh
  vacuum in the far past, then in the exterior Hawking radiation corresponds to a flux
  of positive energy radiation travelling outwards from near the surface to future infinity.
  Based on pair creation, the evaporation of the collapsing star can be equivalently described
  by the absorption of an ingoing negative energy flux of radiation travelling
  towards the center of the star. Here, we are interested in the
  evolution of the star during its collapse. Thus we include the
  backreaction of the negative energy Hawking flux in the interior
  geometry of the collapsing star when writing the full 4-dimensional
  Einstein and hydrodynamical equations. Hawking radiation emitted before the star passes through its
  Schwarzschild radius slows down and reverses the collapse of the star. The star evaporates without forming an horizon or a singularity. 
  This study provides a more realistic investigation than the one
  first presented in \cite{laurabh}, since the
  backreaction of Hawking radiation flux on the collapsing star is studied in the case when the initial state of the field is in Unruh's vacuum.

\end{abstract}


\maketitle

\section{Introduction}
The backreaction of Hawking radiation on the evolution of the collapsing star is the most important problem in the quantum physics of black holes. This problem provides an arena for the interplay of quantum and gravitational effects on black holes and their respective implications for the singularity theorem.
A key feature of Hawking radiation, which was well established in seminal works by ~\cite{unruh, hawkingbh, pc, parkerbh, daviesunruh, bhliterature, candelasunruh, movingmirror}, is that radiation is produced during the collapse stage of the star, prior to black hole formation ~\cite{spindel}. The very last photon making it to future infinity and thus contributing to Hawking radiation, is produced just before an horizon forms. The effect of Hawking radiation on the collapse evolution of the star was for the first time considered in \cite{laurabh}. In \cite{laurabh} it was shown that once the backreaction of Hawking radiation is included in the interior dynamics of the star, then the collapse stops and the star bounces. Solving analytically for the combined system of a collapsing star with the Hawking radiation included, is quite a challenge. The system studied in \cite{laurabh} was idealized in order to obtain an approximate analytical solution: there the star was taken to be homogeneous; the star's fluid considered was dust; the star was placed in a thermal bath of Hawking radiation which arises from the time-symmetric Hartle-Hawking initial conditions on the quantum field in the far past. Within these approximations, the main finding of \cite{laurabh} was that a singularity and an horizon do not form because the star reverses its collapse and bounces at a finite radius due to the balancing pressure of the negative energy Hawking radiation in its interior. Yet, the evolution of the star could not be followed beyond the bounce with the approximate analytic methods of \cite{laurabh}.

Given the fundamental importance of this problem and the intriguing results of \cite{laurabh}, we here aim to study the backreaction of Hawking radiation on the collapsing star by considering a more realistic setting, namely: we allow the star to be inhomogeneous and, based on pair creation near the surface, we consider an Hawking radiation flux of negative energy which propagates in the interior of the star, with its counterpart of positive energy flux travelling fron near the surface outwards to infinity. The flux of Hawking radiation arises when the initial conditions imposed on the quantum field on the background of the star, are chosen to be in the Unruh vacuum state in the far past \cite{unruh, candelasunruh}. In contrast to the Hartle-Hawking initial state which leads to an idealized time symmetric thermal bath of radiation present before and after the collapse, the choice of the Unruh vacuum describes a {\it flux} of thermal radiation which is zero before the collapse and switches on after the collapse. Particle creation occurs near the surface of the star and most of it on the last stages of collapse. Since particles are global objects with ambiguous meaning in curved spacetime, we base our treatment here on the stress energy tensor of radiation that, in contrast to particles, is a local well defined object. The full 4 dimensional set of Einstein and of total energy conservation equations lead to a complete set of hydrodynamic equations for this model. The set of hydrodynamics equations determines the evolution of the star and the absence of singularities.

The paper is organized as follows: Section~\ref{sec:model} describes the metric in the star's interior, the stress energy tensor for the star and for the Hawking radiation flux, which comprise the model we wish to study. We then set up the system of the evolution equations that need to be solved for the combined system and the boundary conditions.  In Sect.~\ref{sec:Results} we describe the evolution of the collapsing star determined by the set of equations in this model. I am indebted to H.Pfeiffer for numerically implementing these equations in his program, shown in the plots in the appendix.  Conclusions for the evolution of the collapsing star when the backreaction of Hawking radiation is included are in Sect.~\ref{sec:conclusions}.



\section{Model}
\label{sec:model}

Let us start with a spherically symmetric and inhomogeneous dust star, described by the following metric

\begin{equation}
ds^2 = - e^{2\Phi(r,t)} dt^2 + e^{\lambda}dr^2 + R^2 d\Omega^2
\label{metric}
\end{equation}
where $R(r,t)$ is the areal radius and $d\Omega^2 = d\theta^2 + \sin(\theta)^2 d\phi^2$.
This form of the metric is convenient for describing a radiating star since due to Hawking radiation we will deal with time dependent, {\it nonvacuum solutions} to Einstein equations. The set of Einstein and energy conservation equations for this metric in the case of an outgoing neutrino flux, were originally derived in Misner~\cite{hmisner}. Our set of hydrodynamic equations is similar to those in ~\cite{hmisner} and \cite{yamada}, but modified to reflect our particular system of ingoing negative energy flux in the interior and positive energy outgoing flux in the exterior of the star.

The changing gravitational potential outside a massive star collapsing into a black hole gives rise to particle creation \cite{pc,parkerbose,movingmirror,bhliterature,spindel, unruh}. A common view of this process at a microphysical level is that the tidal forces of the gravitational field of the star near its surface, rip apart vacuum pairs of particles antiparticles. Of course in this picture we must remember that particles are globally defined object, thus they lose meaning in curved spacetime. For this reason we here consider only the stress energy tensor of Hawking radiation which is a well defined object at each spacetime point.

Based on pair creation, there is an outgoing flux of positive energy particles traveling to future infinity and an ingoing flux of negative energy particles travelling from the surface towards the center of the star. The production of fluxes switches on gradually at the onset of collapse but most of it is produced on the last stages of collapse as the radius of the star is getting close to its horizon value. The particle flux reaching future infinity, measured in the frame of a radially moving observer, becomes Hawking radiation with a luminosity $L_H$. Although most of Hawking radiation is produced on the last
stages of collapse it still occurs before the star crosses the horizon \cite{daviesunruh, pc, unruh, movingmirror,
  spindel}, as can be seen from the expression of Hawking radiation luminosity \cite{pc,parkerbose,movingmirror,bhliterature,spindel, unruh} which goes to zero when taking the limit $R_s -> 2 M$. 


We now wish to include the backreaction of the ingoing negative energy
Hawking radiation flux on the interior of the star and solve for the
star's evolution. The distribution of the stress energy tensor of Hawking radiation in the interior of the star is not yet known.

In our model we assume the stress energy tensor of the Hawking radiation flux in the interior of the star is given by
\begin{equation}
\tau^{ab} = - q_H k^a k^b
\label{hawkingenergy}
\end{equation}

where $k^a$ is an ingoing null vector $k^a k_a = 0$ defined by $ k^a = (e^{-\phi}, - e^{-\lambda/2},0,0 )$ and $q_H$ is defined as the energy density measured locally by an observer with 4-velocity $u^{\mu}$ by $q_{H} = \frac{u^{\mu} u^{\nu} \tau_{\mu\nu}}{8 \pi}$. We explicitly introduce the minus sign in the expression for the negative energy flux in the interior, Eq.\ref{hawkingenergy}, for the simplicity of keeping the parameter $q_H > 0$.

The magnitude of the radiation energy density and flux, $q_H$, in the interior is related to Hawking's radiation luminosity at infinity via the Kodama symmetry \cite{kodama}.
Since our metric is time dependent, there is no Killing vector symmetry $\partial/\partial t$ with respect to which the positive frequency modes for the Unruh vacuum at past infinity can be defined. Instead there is a symmetry related to the Kodama vector $K = e^{-\phi} \partial/\partial t$  with respect to which positive frequencies can be defined and which leads to a conservation law, the conservation of fluxes. The conservation law based on Kodama symmetry, states that the net radiation flux around an imaginary surface with radius $R_0$ enveloping the surface of the star is equal to the net Hawking flux at future infinity, i.e. the mass of the star. The enveloping surface does not neccessarily have to be a trapped surface or an apparent horizon.  The luminosity of radiation at any spacetime point is a variable we solve for, $L = 4 \pi R^2 q_H$, which at infinity we require becomes equal to Hawking radiation $L_H$.

 Since our goal is to study the interior evolution of the star, then we are only interested in the net amount of flux that goes inside the star, rather than outside. In fact a study of the interior and exterior regions which covers the whole spacetime and allows analytic continuation from the interior to the exterior of the star, would require an extended spacetime metric such as the Painleve metric. Although our metric does not cover the whole spacetime, it covers the regions of interest, the interior of the star and part of the exterior to future infinity.


The form of the renormalized stress energy tensor in the exterior of the star, for the Hawking radiation flux in the Unruh vacua, at future infinity and near the surface of the star were calculated in Candelas \cite{candelasunruh}. Based on Kodama's conservation law of Hawking radiation at infinity to that at the star's surface \cite{kodama}, we take the net inward negative energy flux in the star's interior equal and opposite to the net amount gained at infinity, \cite{daviesunruh, candelasunruh, hawkingbh,kodama}. Relating the radiation flux gained at infinity to the radiation flux lost by the star via the Kodama conservation of fluxes, enables us to avoid transplanckian problems that would arise from large blueshifting effects if we had considered a scenario of tracing the quanta of radiation back from infinity to the surface of the star.

We consider the quantum field to be in the Unruh initial vacuum state as it provides a more realistic approximation of the state that follows gravitational collapse. When the initial state of the field is in Unruh's vacuum, there is no flux coming from the far past and before the star collapses. The flux of radiation is produced once the collapse starts until the star reaches the horizon. Then the quantum stress energy tensor of the Hawking radiation flux in the interior, consistent with Hawking radiation flux at future infinity, (with notation $L >0 $), from Eq.~\ref{hawkingenergy} is as follows

\begin{equation}
\tau_{a}{}^{b} = \frac{L}{4 \pi R^2} \left(
\begin{matrix}1 & - e^{\phi - \lambda /2} & 0 & 0 \\
 e^{- \phi + \lambda /2} & -1 & 0 & 0 \\
0 & 0 & 0 & 0 \\
0 & 0 & 0 & 0
\end{matrix}
\right)
\label{candelas}
\end{equation}

The fluid of the inhomogenenous dust star with a 4-velocity $u^a$, normalized such that $u^a u_a = -1$, has a stress energy tensor
\begin{equation}
\label{eq:Matter}
T^{ab} = \epsilon u^a u^b + p (g^{ab} + u^a u^b ),
\end{equation}
where the energy density $\epsilon$ is expressed in terms of a specific internal energy density per baryon $e$ and the number density of baryons $n$ by $\epsilon = n(1+e) =n h$ with $h$ the enthalpy and $u^a = (e^{-\phi},0,0,0)$.
We consider the stellar material to be dust $p=0$ in what follows. In the stellar interior, the total stress energy tensor of the combined system of the star's fluid ($T_{ab}$) and of the radiation ($\tau_{ab}$) is
\begin{equation}
T^{ab}_{total} = T^{ab} + \tau^{ab}.
\end{equation}

The equations that describe the dynamics of the interior of the star with the backreaction of the Hawking radiation flux included are: the Einstein equations,

\begin{equation}
G^{ab} = T_{total}^{ab},
\label{einstein}
\end{equation}

the total energy conservation equations,

\begin{equation}
\label{energyconserve}
\nabla_b T_{total}^{ab} = 0,
\end{equation}

which implies
$\nabla_b T^{ab} = - \nabla_b \tau^{ab}$,
and the baryon number conservation
\begin{equation}
\nabla_a (n u^a) = 0.
\label{baryon}
\end{equation}


The total energy conservation is explicitly written as a set of two equations which contain the radiative heat and momentum transfer between the fluid and the Hawking flux, in our case the absorption by the star of the ingoing negative energy Hawking flux. Since our system transfers both energy and momentum to the star, then the energy conservation is given in terms of a vector $Q^a$, 
with the t-component and the r-component, obtained by contracting $T^{ab}$ in \ref{energyconserve} with $u^a$ and the unit vector $n^a$ orthogonal to the 3-hypersurface $\Sigma_t$ at $t=constant$, as follows
\begin{equation}
Q^a = \nabla_b \tau^{ab} = (e^{-\Phi} nC, e^{-\lambda/2} nC_r, 0,0)
\label{hawkingconserve}
\end{equation}

From Eqn.~\ref{hawkingconserve} and the Hawking flux stress energy tensor in the interior Eqn.~\ref{candelas}, it is straightforward to demonstrate that $Q_t = -Q_r = ( e^{-\lambda}D_{t} (L e^{\lambda}) - e^{-2\phi}D_{r} (L e^{2\phi})\frac{1}{4\pi R^2}  $, with $D_t$ and $D_r$ defined below in Eqn.\ref{eq:ProperDerivatives}.  
We define a 'heat' transfer rate $C$ by $Q_t = n(r,t) C$ and a momentum transfer rate $Q_r = n C_r$ where $ C_r = - C$. The latter simply reflect the fact that the star is receiving negative energy travelling inwards in its interior.

We can now write explicitly the full set of hydrodynamic equations, given above. First let us define the proper derivates $D_r,D_t$ by the following relation in terms of the metric
\begin{equation}
\label{eq:ProperDerivatives}
D_r = e^{-\lambda/2}\partial_r \,\,\,\,
D_t = e^{-\Phi} \partial_t .
\end{equation}

The evolutionary equations for baryon number density $n(r,t)$ from \ref{baryon}, specific
internal energy $e$ from \ref{energyconserve}, the energy of null-radiation with luminosity $L$ given from the t-component of \ref{hawkingconserve} and \ref{energyconserve}, the proper velocity of the comoving fluid $U$, areal radius of comoving fluid elements $R(r,t)$, and the 'acceleration' equation obtained from the r-component of \ref{hawkingconserve} and the Einstein equations \ref{einstein}


\begin{subequations}
\label{eq:partialPDE}
\begin{align}
\label{eq:Dtn}
&\frac{1}{n}D_tn + \frac{1}{R^2}\frac{\partial}{\partial R}(R^2U) = 4\pi R q_H \alpha f^{-1/2},\\
\label{eq:Dte}
&D_t e=-C - p D_t\left(\frac{1}{n}\right),\\
\label{eq:DtL}
& D_tL = -4\pi R^2 n C + e^{-2\Phi}D_r\left(L e^{2\Phi}\right)
 -2L\left(\frac{\partial U}{\partial R}-\frac{L\alpha}{Rf^{1/2}}\right),\\
\label{eq:DtR}
& D_tR=U,  D_r R = f^{1/2}, \\
\label{eq:DtU}
&D_tU = f \frac{\partial\Phi}{\partial R}-\frac{m}{R^2} - 4 \pi R p + \frac{L}{R}.
\end{align}
Two auxiliary quantities are given by radial ODEs,
\begin{align}
\label{eq:dmdR}
  \frac{\partial m}{\partial R}&=4\pi R^2\left(\varepsilon -
    q_H \left(1-\alpha U f^{-1/2}\right)\right),\\
\label{eq:dPhidR}
  (\varepsilon+p)\frac{\partial\Phi}{\partial R}&=-\frac{\partial
    p}{\partial R}+ n C f^{-1/2},
\end{align}
\end{subequations}
and a third one by $D_{t}m = - L ( \sqrt{f} - U) - 4\pi R^2 p$ . We have shown terms that depend on $p$ in the set of equations above for completeness. However in our model we set $p=0$. 
Finally, $\varepsilon$, $f$, $\alpha$ and $R'$ are short-cuts for
\begin{align}
\label{eq:f}
  f &= 1+U^2-\frac{2m}{R},\\
  \varepsilon&=n(1+e),\\
  \alpha& = \frac{R'}{|R'|}=\mbox{sign}{R'},\\
  R'&=\frac{\partial R}{\partial r}.
\end{align}

A useful parameter for tracking the formation of the horizon and the singularity is the expansion parameter $\theta_{(K)}$ of $r=$const surfaces along
the outgoing null-normal
$K^\mu = D_t + D_r = e^{-\Phi}\partial_t + e^{-\lambda/2}\partial_r$, given by
\begin{equation}
\label{eq:theta}
  \theta_{(K)} = \frac{2}{R}\left(\alpha\sqrt{f} + U\right).
\end{equation}
When $\theta_{(K)}=0$, an apparent horizon has formed.

\subsection{Boundary Conditions}

We need to determine the surface luminosity of the star measured in the local frame of a radially moving observer in the interior, located at $R= R_S$ with velocity $U_S$. This problem for the case of a star emitting a positive energy null flux was studied in \cite{hmisner}. There, an observer time at infinity $u_{\infty}$ of an event was defined as the time which an outgoing radial light ray emitted from the event reaches a distant observer at future infinity. Our problem is slightly more complicated in the sense that we have two fluxes which are produced not by the stellar material but rather by the changing gravitational field, a.k.a the curved spacetime, near the surface of the star. In the interior we have an ingoing negative energy flux through the surface towards the center of the star. In the exterior of the star we have an outgoing positive energy null radiation from the fuzzy region near the surface of the star, where particle creation occurs, to future infinity. In this part of spacetime our problem is identical to that studied in \cite{hmisner}. We can make use of the derivation of luminosity as a function of coordinates in the exterior of the star, of \cite{hmisner} sketched briefly below.
 Outgoing and ingoing light rays for our metric, are determined by the condition $e^{\phi} dt = \pm e^{\lambda/2} dr$ with $(+)$ and $(-)$ for outgoing and ingoing respectively. For the exterior metric a retarded time {\it u}, which far away from the radiating star becomes $u_{\infty}$, is defined by
\begin{equation}
 e^{\psi} du = e^{\phi} dt - e^{\lambda/2} dr
\end{equation}
where $e^{\psi}$ is an integrating factor to make $du$ an exact differential. Then $u=constant$ are null hypersurfaces that contain outgoing null radiation.
Our metric Eqn. ~\ref{metric}, rewritten in terms of the 'observer' time $u$ in the exterior containing the outgoing positive energy flux, becomes a Vaidya metric, (see \cite{hmisner} for the details)

\begin{equation}
ds^2 = - e^{2\psi} du^2 - 2 e^{\psi} e^{\lambda/2} du dr + R^2 d\Omega^2
\end{equation}

 Let us imagine an imaginary surface close to the surface enveloping the star at $ R \simeq R_S$ in the exterior region that contains the outgoing null radiation. Then,
\begin{equation}
D_t = e^{-\psi} \frac{\partial}{\partial u}.
\label{derivativeu}
\end{equation}
and on this surface close to the surface of the star $\psi$ satisfies: $ e^{\psi} \simeq (\sqrt{f_s} + U_s)$ .
Outside the star, the only nonvanishing component of Ricci tensor is $R_{uu} = -\frac{2}{R^2} \frac{dm}{du}$ which together with Einstein equations $D_{t} m = - L (\sqrt{f} + U)$ and $\frac{u^{\mu} u^{\nu} R_{\mu\nu}}{8 \pi} = q_H$, determines the luminosity $L(r,t) = 4 \pi R^2 q_H$ of the positive energy radiation in the exterior region at each point in the local frame of a radially moving observer with 4-velocity $u^\mu$. The luminosity of Hawking radiation at future infinity ($R -> infty , U -> 0$) is given by 
$L_H = - \frac{dm}{du_{\infty}}$.

 Using the relation of Eqn.~\ref{derivativeu} and the definition of Hawking luminosity at infinity $L_H$ results in a luminosity relation for the outgoing flux at location $(r,t)$ in the exterior
\begin{equation}
 L(r,t) = \frac{L_H}{(\sqrt{f(r,t)} + U(r,t))^2} ,
\end{equation} 
which in the fuzzy region near the surface of the star $R\simeq R_S$ yields  $L_S = \frac{L_H}{(\sqrt{f_S} + U_S )^2}$  ~\cite{hmisner}. 

But the interior of the star, with the metric given by Eq.1 matched to Vaidya metric in the exterior, contains an ingoing negative energy flux of null radiation. The Einstein equation for the mass loss with time due to the absorption of negative energy radiation in the interior is $D_{t}m(r,t) = -L(r,t) (\sqrt{f} -U)$ where $(\sqrt{f(r,t)} - U(r,t) )$ is the redshift factor of energy, (also the time dilation factor of energy absorbed), at each spacetime point $(r,t)$ in the interior.
A radially moving observer in the interior positioned at the surface of the star $R=R_S$ , with 4-velocity $u_\mu$ , measures an energy density and luminosity

\begin{equation}
L_{H,S} = u_{\mu} L_S k^{\mu} = L_S (\sqrt{f_S} - U_S).
\label{surfacelum}
\end{equation}   
 $L_{H,S}$ provides the boundary condition for the luminosity of the ingoing radiation from the surface towards the center in the interior of the star.
The relation between $dt$ for the interior metric and observer time at infinity $du_{\infty}$ can be obtained from Einstein equations and the continuation of the interior Eqn. \ref{metric} to the exterior Vaidya metric at the surface, (using Eqn. \ref{derivativeu} and the relations: $dm/du_{\infty} = (dm/dt)(dt/du_{\infty})= - L_H$.

{\it The energy absorption rate \boldmath$C$:} The quantity $C(r,t)$ determines
the absorption rate. It is such that
\begin{align}
\label{eq:mBC}
m&=0,  L=0, \qquad\mbox{at $r=0$, for all $t$},\\
\label{eq:PhiBC}
\Phi&=0\qquad\mbox{at $r=r_{\rm star}$, for all $t$}.
\end{align}

In the stellar interior, the ingoing radiation $L$ is absorbed by the
stellar material.  This reduces $L$ (i.e. the magnitude of the
radiation), and since the ingoing flux carries negative energy, it
also reduces the mass of the star. Again, we choose to display negative signs
explicitly in Eqns.~(\ref{eq:Dte}) and~(\ref{eq:DtL}), so that a
positive $C$ corresponds to a reduction in stellar mass (through
driving $e$ more negative), and in a reduction of $L$.

We demand that the ingoing negative energy Hawking flux is completely absorbed by the star.
The physical reasons for the expectation that the negative energy Hawking flux must be absorbed completely by the star are two fold: $(i)$ if part of the negative energy flux were not absorbed in the interior before reaching the center, then it could bounce off the center $r=0$ and travel outwards, in which case the negative energy flux would disturb the thermal nature of radiation and it could travel all the way to future infinity; $(ii)$ if the negative energy ingoing flux were not completely absorbed in the interior that would also imply that the star is not evaporating as fast and it is violating the mass loss relation $dM/dt \simeq 1/M^2$ expected of evaporation by Hawking radiation.Either one of these options would result in a nonconservation of fluxes and in a radiation by the star which is not consistent with Hawking radiation.
$C(r,t)$ then has to be chosen such that the entire ingoing flux $L$ is absorbed approximately
uniformly through the star, to achieve an approximately uniform
evaporation of the stellar material, (a particular example of such $C(r,t)$ is shown in the Appendix).

\section{The evolution of the collapsing star}
\label{sec:Results}

The set of Eqns.~(\ref{eq:partialPDE}) determines the evolution of the interior for the gravitationally collapsing star. These equations were numerically programed by H. Pfeiffer. The results are shown in the plots in the Appendix.
Stars with different initial radius $R_0$ and with
different initial masses $M_0$ were investigated.  To reduce the size of
  parameter space, the initial compactness is chosen to be
  $R_0/M_0=10$.  

Figure~\ref{fig:033_M4_Overview} shows a
 typical outcome, a star with initial mass $M_0=4$.  The evolution
proceeds initially through a phase which is almost identical to the standard
Oppenheimer-Snyder collapse. The similarities with the Oppenheimer-Snyder collapse end as the star approaches its
Schwarzschild-radius, which would happen when $R_S=2m_S$.  During the last stage of collapse, before
the star crosses its Schwarzschild-radius, the luminosity $L_{H,S}$
increases rapidly, resulting in substantial reduction of the total
mass $m_S$.  The mass $m_S$ reduces such that $2m_S$ remains 
smaller than $R_S$. In all cases investigated, for a variety of initial mass $M$ and initial radius $R$, the same results persist: the star does not cross the horizon and a singularity does not form, as can be seen by $\theta_{K} >0$ in Fig.3. Instead a reversal of collapse is observed and the star bounces as it gets close to the horizon (Fig.2 and Fig.3).
 The numerical program failed to converge when the mass dropps half of its initial value. Therefore the evolution of the star after it reverses its collapse and what is left behind, is not known and may require more sophisticated numerical techniques.


 Quantities such as $L_{H,S}$, $R_S-2m_S$ and the expansion
$R/2\theta_{(K)}$) change rapidly during the transition, and then
settle down to almost constant values. This behaviour can be seen in Fig.~\ref{fig:033_M4_Zoom} where the transition from quasi
Oppenheimer-Snyder collapse to the evaporation phase is enlarged. As can be seen, the parameter $\theta_{K} >0$ at all times, which indicates that horizons and singularities are absent in this model.

The transition from the quasi Oppenheimer-Snyder collapse to the
evaporation phase is universal, in that many of its features are
independent of the mass $M_0$.  The evaporation phase shows several
universal features. An intriguing universal feature displayed by the stars, is that on the last stages before the collapse stops, they reach Planck luminosity, the maximum luminosity a star can have, independently of their mass and radius, while having the correct luminosity for Hawking radiation at infinity. The universal features can be seen in Fig.~\ref{fig:L_032_033_034} for $L_{H,S}$ and
$R_S-2m_S$ for simulations with a variety of initial masses $M_0=4,8,16$.

\section{Conclusions}
\label{sec:conclusions}

Einstein equations tell us that the final destiny of a gravitationally
collapsing massive star is a black hole \cite{os}. This system
satisfies all the conditions of the Penrose Hawking singularity
theorem \cite{singularity}. However a collapsing star has a spacetime
dependent gravitational field which by the theory of quantum fields on
curved spacetime should give rise to a flux of particles created
\cite{pc}. Hawking discovered in the early $'70's$ that this is
indeed the case: stars collapsing to a black hole produce
Hawking radiation \cite{hawkingbh}. The conclusions derived from both
theories, the existence of black holes from Einstein's theory of
gravity and the existence of Hawking radiation from the theory of
quantum fields in curved spacetime, were soon found to be in high
friction with one another, (see \cite{ellis} for an interesting
treatment).They led to a series of paradoxes, most notably the
information loss \cite{infoloss}. Being
forced to give up on either unitarity or causality is at the heart of
this longstanding problem. Quantum theory is violated if unitarity is
broken. Einstein's theory is violated if causality is
broken. Violations of the quantum theory imply Hawking radiation may
not exist.  Violations of Einstein's theory of gravity, on which the
singularity theorem is based, imply black holes may not exist. Thus
black hole physics provides the best arena for understanding the
friction between quantum and gravitational physics. In this light, an
investigation of the evolution of the star's interior as it is
approaching its singularity, with the backreaction of the quantum Hawking
radiation included, is of fundamental importance and it may offer a way out of this conundrum.

This problem was first investigated with semianalytical methods and a series of approximations, such as a homogeneous star placed in the Hartle Hawking thermal bath in \cite{laurabh}.

In this work we extend our investigation to study a more realistic
system of a collapsing star with the quantum field in the Unruh
vacuum, which gives rise to a Hawking flux produced only during the
collapse. Here we consider the collapsing star absorbing an ingoing negative energy Hawking flux. The net amount of flux was considered approximately equal to the flux that would be produced by the collapsing star if the star had reached the horizon. Based on Kodama's conservation of fluxes we related the net negative energy flux entering from the surface into the interior of the star, to the net Hawking radiation flux at infinity, ensuring that the amount of negative energy absorbed by the star does not exceed the mass of the star.
          
The full set of Einstein and energy
conservation equations for the simple model presented here, determines that the evolution proceeds through two phases:
First, the start of a collapsing phase where Hawking radiation is unimportant, and
the star follows very closely standard Oppenheimer-Snyder collapse.
When the star approaches formation of an horizon, then Hawking
radiation sets in.  This slows down the collapse while significantly reducing the
mass of the star.  Both effects (slowdown and mass-loss) reverse the collapse and the evaporating star remains outside its event
horizon, as indicated by the parameter $\theta_{K} >0$.

The key idea that enables this program is the fact that radiation is produced during the collapse stage of the star, prior to black hole formation \cite{bhliterature, spindel}. Once the star becomes a black hole with a singularity and an event horizon, then nothing can escape it, not even light. The issue of whether Hawking radiation is produced before or after a black hole forms is still debated. The results shown here are based on the view that Hawking radiation is produced during the collapse of the star, with most of it on the last stages of collapse, just before the star crosses the horizon and before a singularity forms. In the latter case the backreaction of Hawking radiation on the star's evolution during collapse towards a black hole, can be included in the set of hydrodynamic equations for the coupled system of the quantum field and the star and it can modify the evolution of the collapse, as we showed here.

The set of hydrodynamics equations is a closed set that suffices to describe the evolution of collapse for the inhomogeneous dust star absorbing negative energy Hawking flux during its collapse. We discover that instead of collapsing to a black hole, the gravitationally collapsing stars do not form a singularity and do not cross what would have been the apparent horizon. Instead they evaporate away. It is interesting that besides the universality of Hawking radiation, the behaviour of the collapsing stars reversing collapse and evaporating away without ever crossing the horizon (and, without a singularity forming), seems also {\it universal}, i.e independent of their characteristics such as mass and size. 

Physically the backreaction of ingoing negative energy Hawking radiation reduces the gravitational binding energy in the star with the maximum loss near the last stages of collapse, while reducing momentum of the stellar material. These two factors are the reason for the reversal of collapse and the absence of singularity and horizon formation. Independent of what size and mass the star starts from, most if its radiation will be produced as the star nears its future horizon. At that stage the drop in mass and internal energy is maximum and the star never reaches an apparent horizon. The ingoing negative energy Hawking radiation absorbed by the star, {\it violates the energy condition of the singularity theorem \cite{singularity}}.Then it is not surprising that a singularity and an horizon do not form, features traditionally associated with the definition of black holes. Stated simply our findings indicate that singularities and horizon do not form due to quantum effects. Universally these black hole candidates evaporate away without crossing the Schwarzschild radius in this model, although from far away they do not appear much different from a black hole due to time dilation effects.

A more accurate study should include the spacetime dependence of the stress energy tensor of the ingoing flux in the interior of the star. The calculation of the radiation stress energy tensor in the interior of the star is left for future work. Here we assumed that the flux enters the star through the surface and propagates inwards at the speed of light. Since the distribution of the stress energy tensor of radiation is not known in the interior, then one could perhaps speculate and think of the star as a collection of massive shells  all obeying relations of the type $dm(r,t)/dt \simeq 1/m^2$.  In this picture, then one could argue that the inner shells would evaporate faster then the outer ones, and 
the star may likely have 'unfolded' from inside. The latter would result in nonlinear dynamics, likely with shell crossings and turbulences. It is possible that if that were to happen and the stress energy of radiation in the interior were known, then the results found here for the reversal of collapse of a simple model of a spherical dust ball with the negative energy ingoing Hawking flux in its interior, may not hold. 

For the model presented here we have shown that due to the backreaction of radiation onto the star, the star does not form  a singularity or an apparent horizon, therefore
 neither unitarity nor causality
are violated. Our findings offer a solution to 
the longstanding information loss
paradox.  Interestingly here universally
 these stars shine with Planck luminosity independent of their mass on their last stage of collapse as they approach the horizon, and deccelerate and reverse the collapse before crossing the horizon. Due to huge time dilation effects,
the evaporation of these objects from faraway would seem so slow that in fact these stars would not look much different from a black hole to a faraway observer. In fact there is a major difference between them and the black holes.
Unlike classical black holes, these 'quantum stars' do not contain singularities.

\acknowledgments
I am grateful to DAMTP Cambridge University for their hospitality when this work was done. Many thanks to M.J.~Perry, L.~Parker, J.~Bekenstein, P.~Spindel, and G.~Ellis for useful discussions, and to D.~Page, W.~Unruh and J.~Bardeen for thoroughly scrutinizing and debating every detail, although we still disagree on the origin of Hawking radiation. I am especially indebted to H.~Pfeiffer for numerically implementing the equations of the model presented here.

\appendix
\section{Numerical Simulations}
\label{sec:numerics}

The numerical simulations of Eqns.(11) shown in the plots in this Appendix, were implemented by H. Pfeiffer.
\begin{figure}
\includegraphics[scale=0.58]{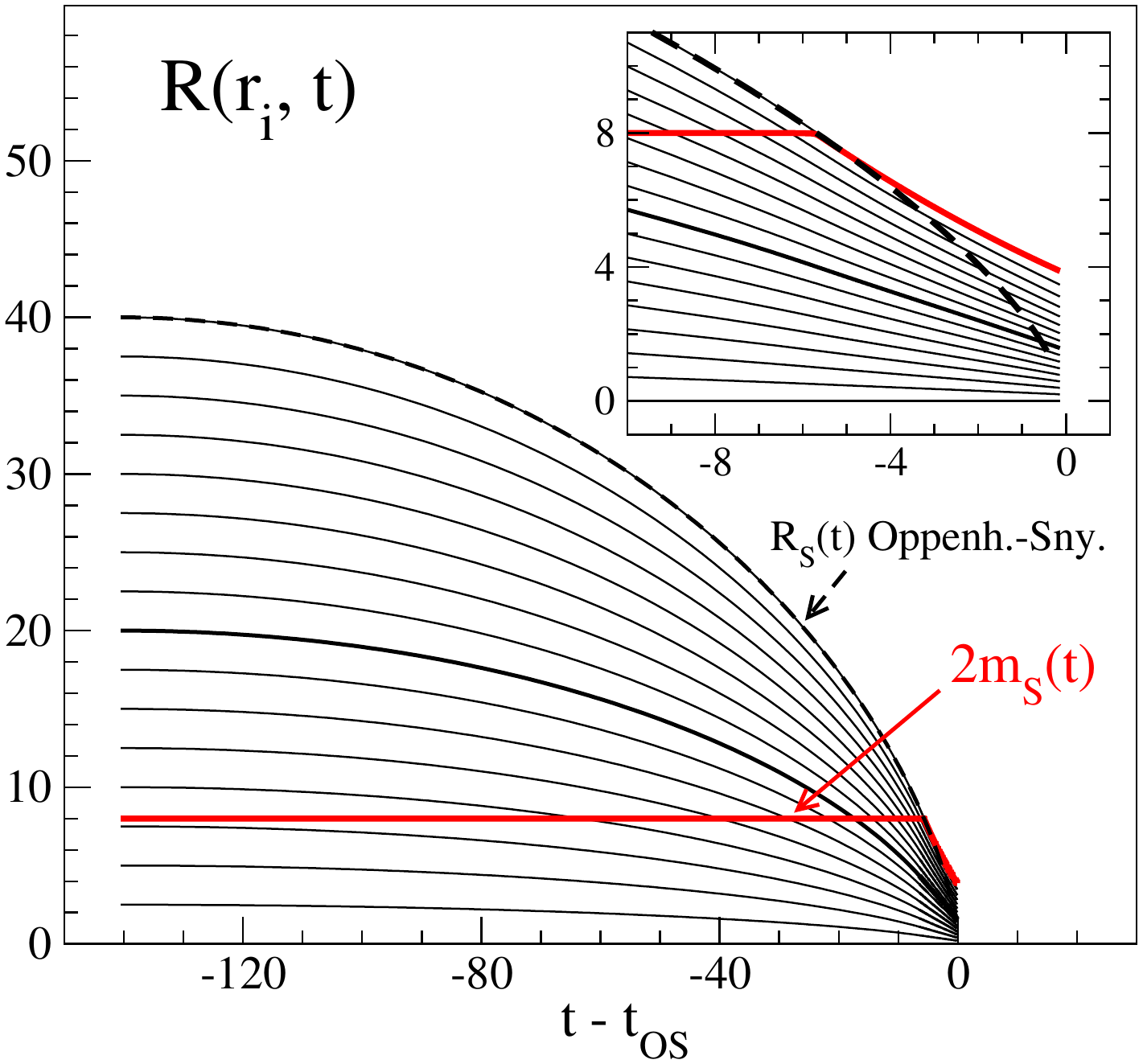}
\caption{\label{fig:033_M4_Overview} Areal radius during the
    collapse with Hawking radiation ($R_0=40$, $M_0=4$).  The black
    lines show areal radii at 16 equally spaced comoving radii
    vs. time ($r=r_S/2$ is indicated with a thicker line).  The black
    dashed curve shows $R_S(t)$ for a standard Oppenheimer-Snyder
    collapse without Hawking back-reaction.  The evolutions proceed
    very similarly until $R_S\approx 2m_S$ (cf. red curve).
    The inset shows an enlargement of late time.  }
\end{figure}
\begin{figure}
\includegraphics[scale=0.58]{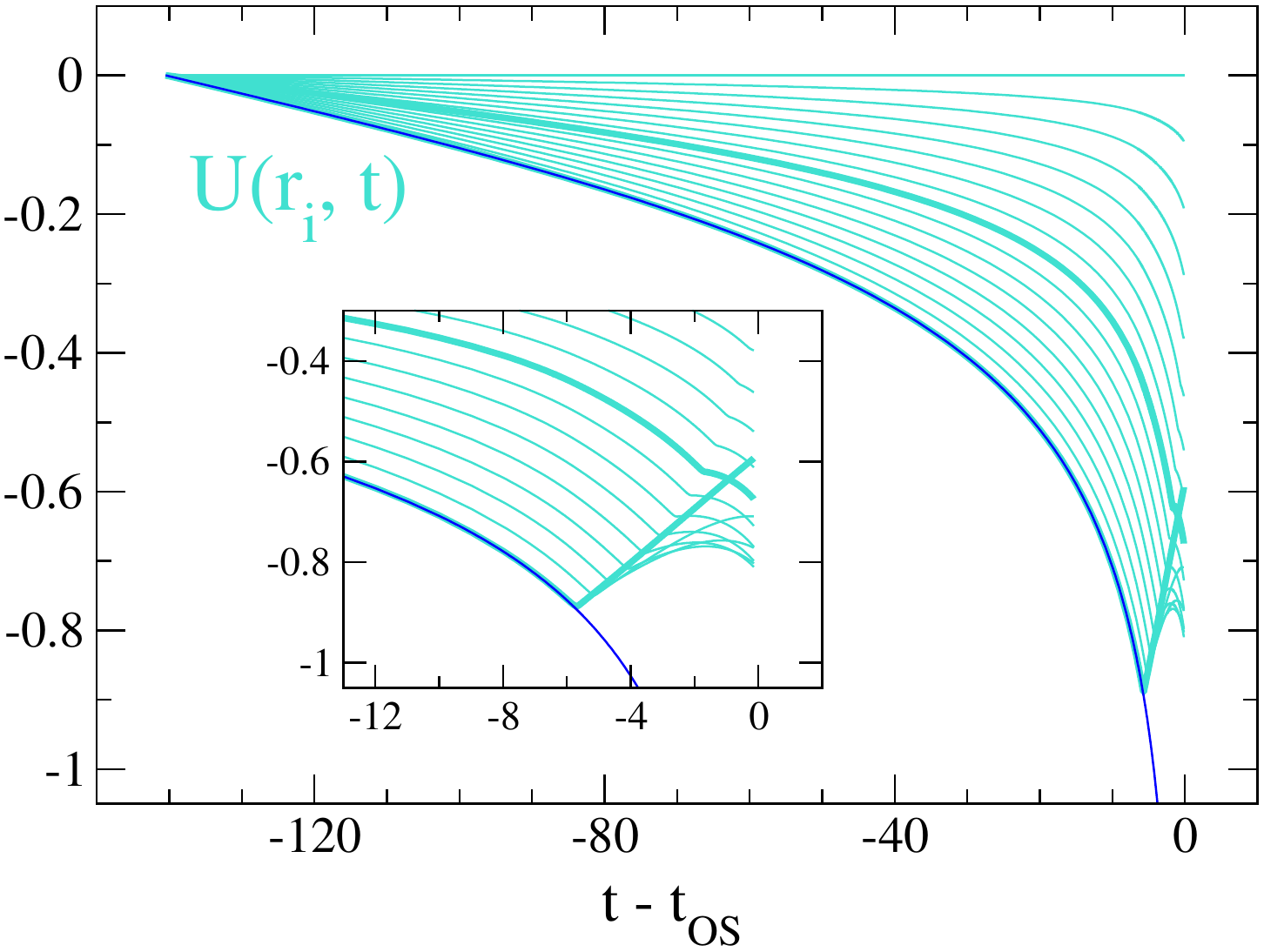}
\caption{\label{fig:033_M4_Overview} Radial velocity $U$ for
    the case $R_0=40$, $M_0=4$.  Shown are the area radii at sixteen
    equally-spaced comoving radii $r_i$, with the thick lines
    indicating the stellar surface $r=r_S$ and the radius $r=r_S/2$.
    The inset shows an enlargement of late time.  The thin blue line indicates $R_S(t)$ for standard Oppenheimer-Snyder collapse. }
\end{figure}
\begin{figure}
\includegraphics[scale=0.5]{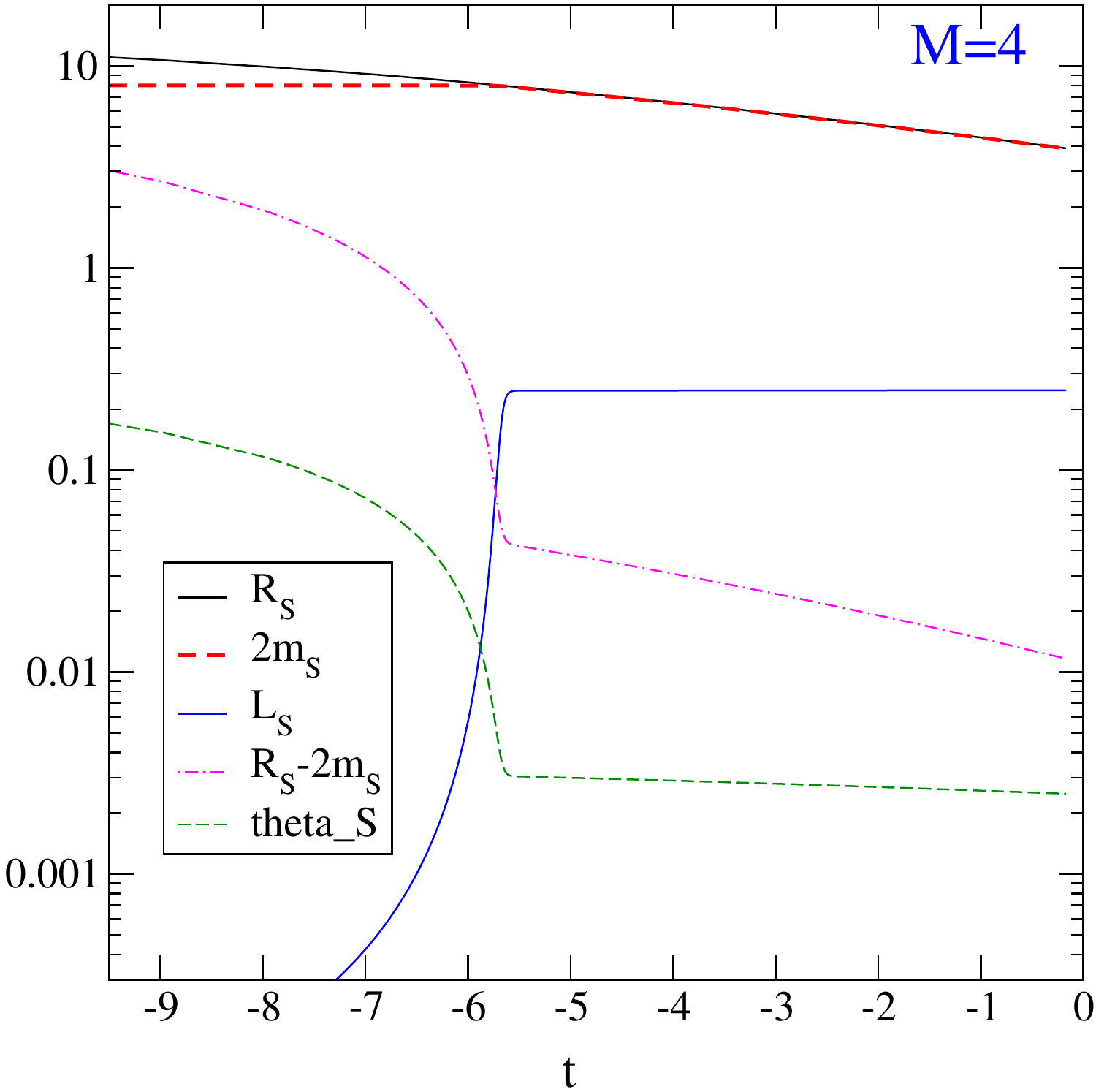}
\caption{\label{fig:033_M4_Zoom} Same evolution as in Fig.~\ref{fig:033_M4_Overview}, but zoomed into the onset of evaporation.}
\end{figure}
\begin{figure}
\includegraphics[scale=0.54]{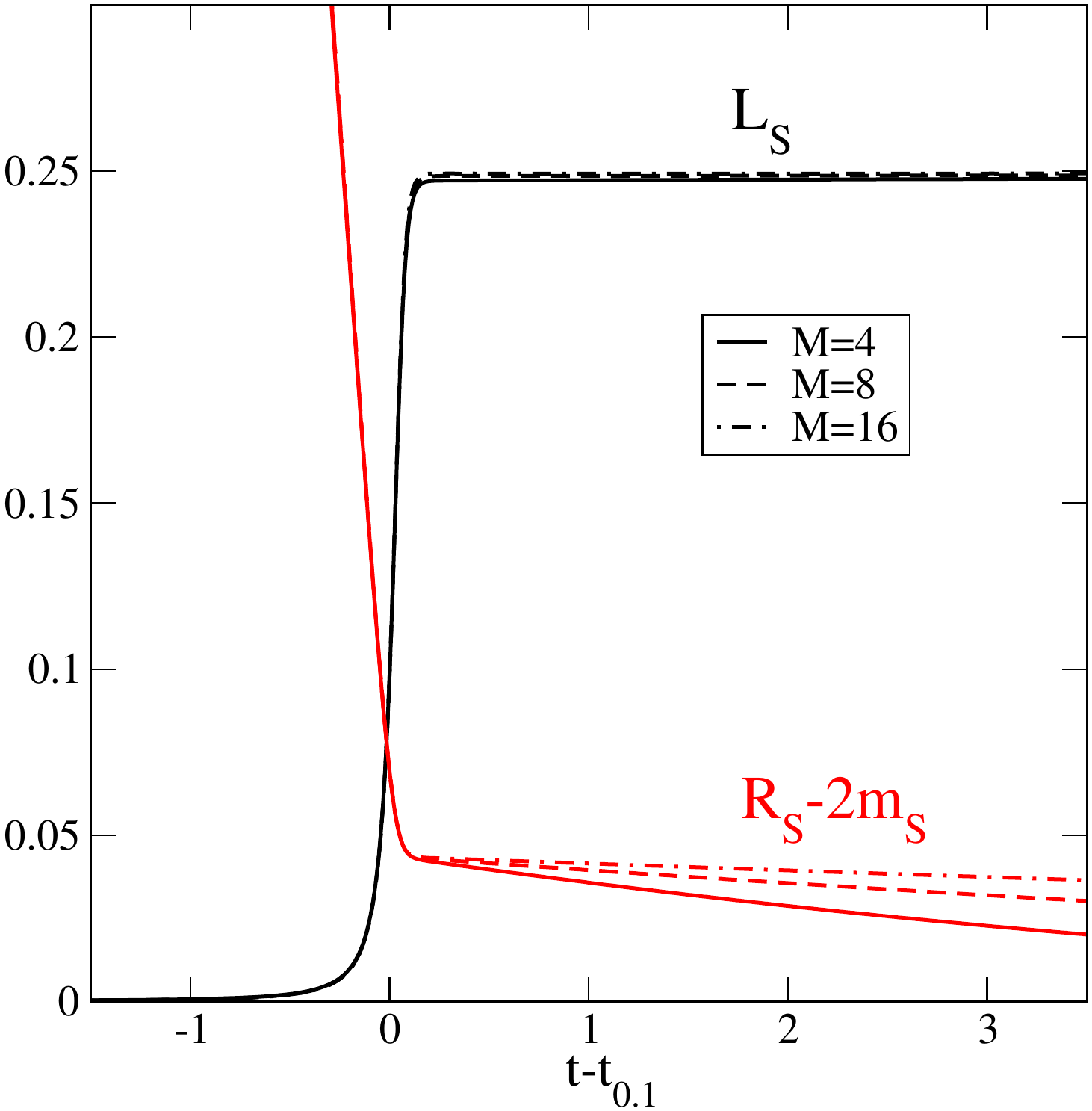}
\caption{\label{fig:L_032_033_034} Universality of the evaporation
  phase and the transition into it.  Shown are simulations for three
  different masses $M_0=4, 8, 16$, with initial radii $R_0=10M_0$.
  The time-axis is shifted separately for each simulation such that
  $L_S=0.1$ for all simulations at $t-t_{0.1}=0$.}
\end{figure}

The numerical implementation uses coordinates $t$ and $r$, where $t$
represents the proper time of the fluid-element on the surface of the
star, with the boundary
condition~(\ref{eq:PhiBC}), and $r$ is the radial coordinate comoving
with the fluid. Eqs.~(\ref{eq:partialPDE}) for this model are implemented in a finite-difference code,
using the Crank-Nicholson method.

 The concrete choice of $C(r, t)$ is motivated by considering the
characteristics of the $L$-evolution equation~(\ref{eq:DtL}).  These
characteristics are ingoing null normals $k^a$,
\begin{equation}\label{eq:l}
k = D_t - D_r=e^{-\Phi}\, \partial_t - \sqrt{f}\alpha\, \partial_R.
\end{equation}
Along the direction of $k^a$, Eq.~(\ref{eq:DtL}) simplifies to an
ordinary differential equation,
\begin{equation}\label{eq:Lalong_l}
k^a\partial_a L = -4\pi R^2 n C + L Q,
\end{equation}
where we introduced the abbreviation
\begin{equation}
  Q\equiv 2 \left( \sqrt{f}\alpha \frac{\partial\Phi}{\partial R}
-\frac{\partial U}{\partial R} + \frac{L\alpha}{R\sqrt{f}}\right).
\end{equation}
We consider propagation along $k^a$ by a small amount $\delta$.  Equation~(\ref{eq:Lalong_l}) shows that the energy flux will change by
\begin{equation}\Delta L = \left(-4\pi R^2n C + LQ\right)\delta,
\end{equation}
whereas Eq.~(\ref{eq:l}) indicates that $R$ will change by
\begin{equation}
\Delta R = - \sqrt{f}\alpha \delta.
\end{equation}
To achieve $L\propto R^3$, we require $\Delta L/L = 3\Delta R/R$, from
which it follows that
\begin{equation}\label{eq:C}
C = \frac{3\alpha\sqrt{f} L }{4\pi R^3 n} + \frac{L Q}{4\pi R^2 n e^\Phi}.
\end{equation}
Equation~(\ref{eq:C}) gives the space- and time-dependent function $C(r,t)$, which is substituted into Eqs.~(\ref{eq:partialPDE}).

\newpage


\end{document}